\documentclass{kapproc}

\usepackage{procps}

\usepackage[dvips]{graphicx}

\upperandlowercase

\let\footnote\savefootnote

\kluwerbib

\let\lcitebracket(
\let\rcitebracket)

\begin{document}

\articletitle{QSO Photometric Redshift estimation for the XMM-Newton/2dF Survey}

\author{S. Kitsionas$^{1}$, E. Hatziminaoglou$^{2}$, I. Georgantopoulos$^{1}$, A. Georgakakis$^{1}$, O. Giannakis$^{1}$}
\affil{$^{1}$Institute of Astronomy \& Astrophysics, National Observatory of Athens, Greece\\
$^{2}$Instituto de Astrofisica de Canarias, Tenerife, Spain}
\email{kitsiona@astro.noa.gr}

\section{Introduction}

\noindent The technique of estimating galaxy and/or QSO redshifts using photometric rather than spectroscopic observations has recently received great attention due to its simplicity and the accuracy of the results obtained.

In this work, we estimate photometric redshifts for a sample of X-ray selected QSOs. This is the first time this technique is applied on such a sample. We first calculate the accuracy of the results obtained by comparing photometric to spectroscopic redshifts for a sub-sample of our QSO sample: for the majority of the objects in this sub-sample ($\sim$67\%) photometric redshift estimates are correct within $\Delta z<0.3$. We then calculate the photometric redshift distribution for the whole QSO sample.

\section{Data}

\noindent We have compiled a sample of 134 X-ray sources from 5 XMM fields at the North Galactic Pole having exposure times from 2 to 10 ksec corresponding to a flux limit of $f($0.5-8 keV$)$=10$^{-14}$ cgs.

The photometry in 5 optical bands and the morphological identification for all members in this sample were obtained from the Early Data Release (EDR) of SDSS. From the 134 objects in our sample, 59 are extended and 75 are point-like. From the 75 point-like objects, 67 are QSOs and 8 are stars based on the flux ratio $f_{x}/f_{opt}$ which is $\ll$-1 for stars. We have obtained spectroscopic redshifts for a sub-sample of 37 QSOs and 18 extended sources from SDSS, the 2dF/GRS, 2QZ, ROSAT and our own observations using the 2dF at AAT.

\section{Results: Photometric {\em vs}. spectroscopic redshifts}

\begin{figure}
\resizebox{3.905cm}{!}{\includegraphics{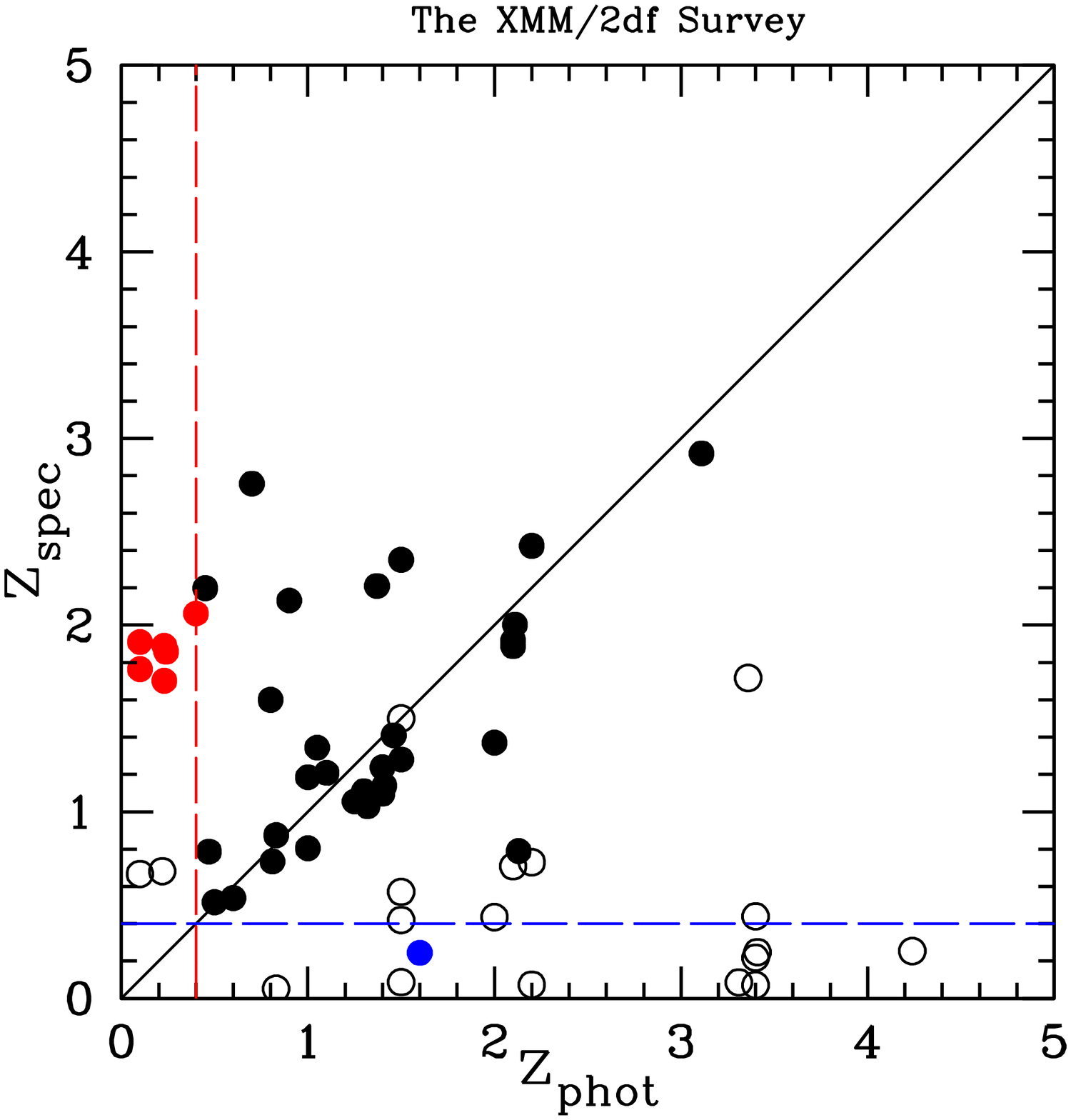}}
\resizebox{3.905cm}{!}{\includegraphics{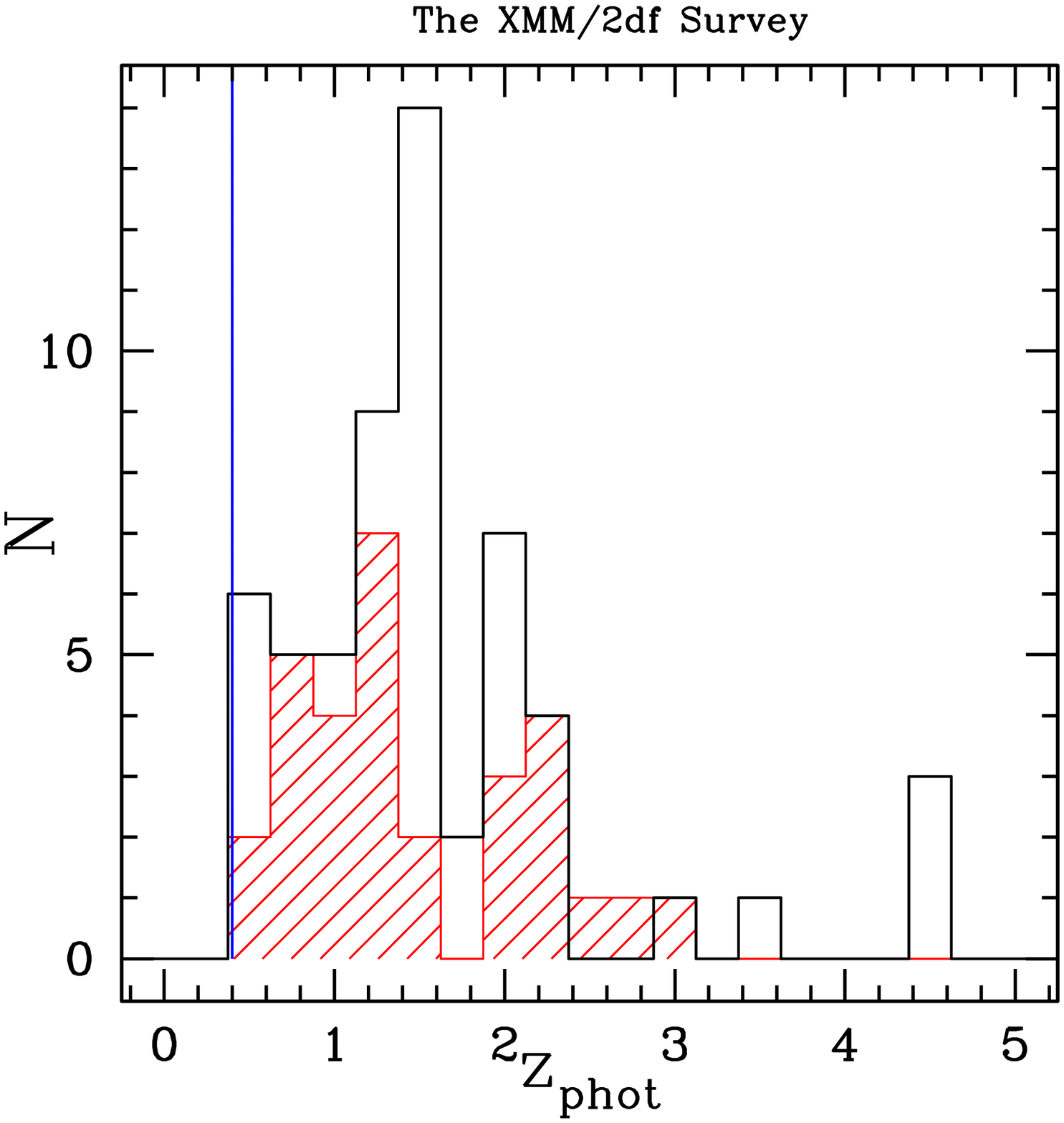}}
\resizebox{3.905cm}{!}{\includegraphics{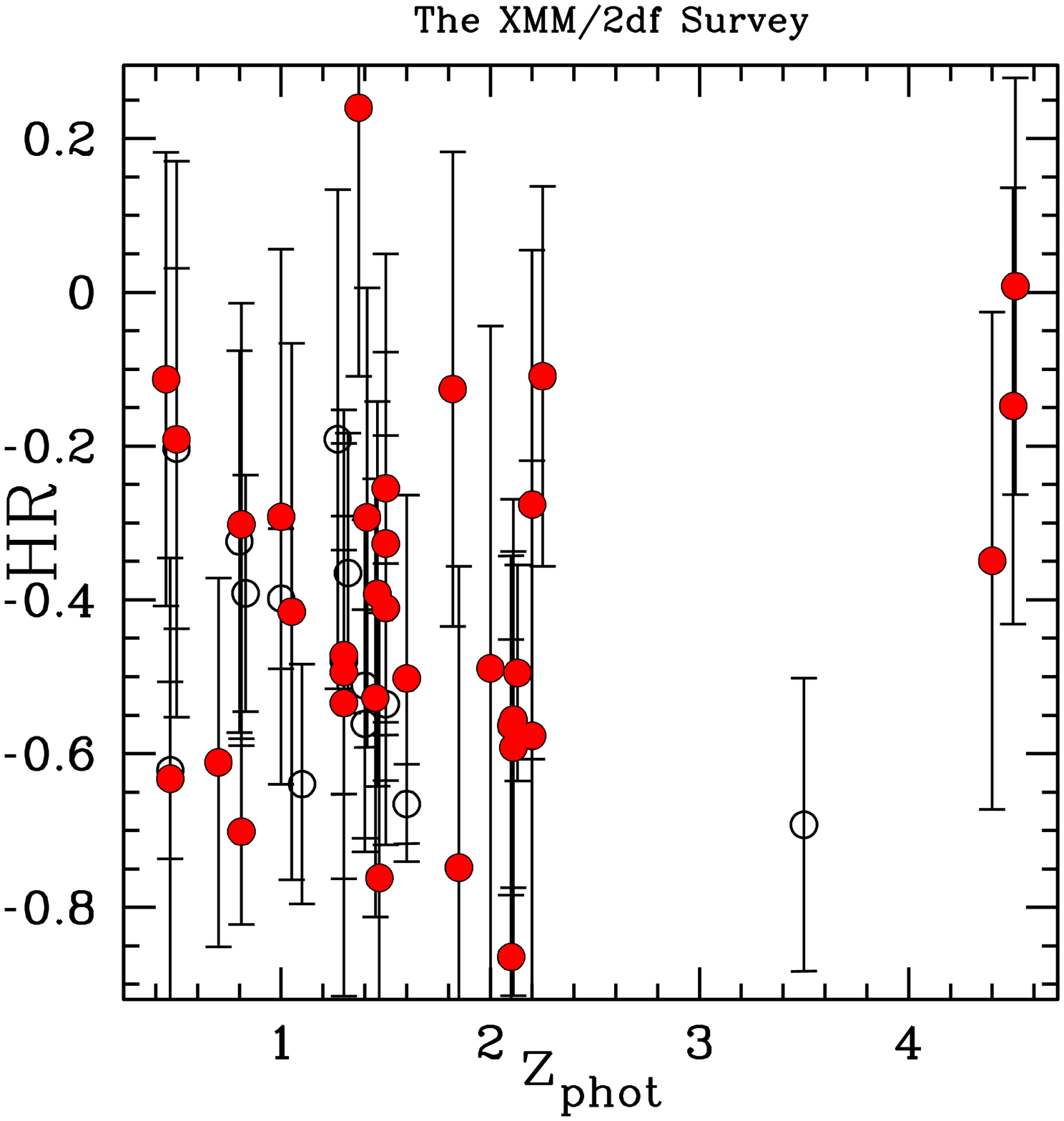}}
\caption{{\em Left}: Comparison between photometric and spectroscopic redshifts. {\em Middle}: Redshift distributions for the whole photometric QSO sample and the spectroscopic sub-sample (shaded). {\em Right}: Hardness ratio {\em vs}. z$_{phot}$ for all objects with 1$\sigma$ errorbars.}
\end{figure}

\noindent We use the algorithm of Hatziminaoglou {\it et al}. (2000) for the simultaneous calculation of photometric redshifts for all sources in the sample. The left panel of Fig. 1 illustrates the comparison of photometric to spectroscopic redshifts for the sub-sample with spectroscopic redshifts. The filled circles are QSOs and the open circles are extended (galactic) sources. It is evident that the method cannot treat (extended) galactic sources as no such templates have been used.

From the 37 QSOs in the spectroscopic sub-sample we have concentrated only to those with $z>0.4$ as our sample indicates that it is very improbable to obtain point-like sources with redshift less than 0.4. For the majority (66.7\%) of the remaining 30 QSOs in the sub-sample, photometric redshifts are correct within $\Delta z<0.3$.

\section{Results: The redshift distributions}

\noindent The middle panel of Fig. 1 illustrates the photometric redshift distribution for the whole QSO sample with $z_{phot}>0.4$ (57 objects). The shaded histogram gives the distribution of spectroscopic redshifts for the 30 QSOs with $z>0.4$ ({\it cf}. the left panel). Statistical tests have shown that the two histograms have been drawn from the same population of objects. This indicates that there is no systematic bias introduced by the photometric redshift estimation technique.

In the future, we expect to use the photometric redshift estimates for all objects to correct for their distance in order to be able to derive the distributions of properties such as the Hardness Ratio and hence the hydrogen column density, the luminosity function for the whole sample etc. As an example, in the right panel of Fig. 1 the Hardness Ratio ($HR$) is plotted against the photometric redshift. We can conclude that there are objects with exceptionally high absorption (one order of magnitude higher than normal) at all redshifts.

\end{document}